\documentstyle[12pt]{article}
\newcommand{\be}{\begin {equation}}
\newcommand{\ee}{\end {equation}}
\newcommand{\ba}{\begin {eqnarray}}
\newcommand{\ea}{\end {eqnarray}}
\begin{document}
\begin{center}
{\Large Relativistic quantum model of confinement and the current
quark masses}
\vspace{0.2cm}

{\large L.D. Soloviev}
\vspace{0.2cm}

{\it Department of Physics, University of Michigan, Ann Arbor, 
MI 48109-1120, USA,\\
  and\\
Institute for High Energy Physics, 142284, Protvino, Russia}
\date{3 April 1997}
\end{center}
\vspace{0.5cm}

\bf Abstract. 
We consider a relativistic quantum model of confined
massive spinning quarks and antiquarks which describes leading Regge 
trajectories of mesons. The quarks are described by the Dirac equations
and the gluon contribution is approximated by the Nambu-Goto straight-line
string. The string tension and the current quark masses are the main
parameters of the model. Additional parameters are phenomenological constants
which approximate nonstring short-range contributions. Comparison of the 
measured meson masses with the model predictions allows one  to determine the 
current quark masses (in MeV) to be 
$m_s = 227 \pm 5,~ m_c = 1440 \pm 10,~ m_b = 4715 \pm 20$. The chiral
$SU_3$ model[23] makes it possible to estimate from here the $u$- and $d$-quark
masses to be $m_u = 6.2 \pm 0.2$~ Mev and $m_d = 11.1 \pm 0.4$~Mev.
 
\vspace{0.5cm}
{\it PACS Numbers: 12.15.F, 12.50.C, 11.17}
\vspace{0.5cm}

It has been believed for a long time that properties of quarks confined in a
meson are closely related to those of the relativistic string with Nambu-Goto 
self-interaction [1]. The anomaly in the quantum string theory in
4-dimensional space-time has led to other important applications of the 
string theory [1]. Nevertheless, the hadron theory can 
use   particular simple configurations of the string for approximate
description of the hadrons if these configurations admit relativistic 
quantization. If the approximate hadron model obtained in this way appears to
be in acceptable agreement with the experiment one can try next more 
complicated string 
configuration, having in mind that at some step the whole notion of string may
fail, especially when more experimental information about hadron daughter
trajectories will be available.

The simplest string configuration, a straight-line string, was quantized in 
[2,3] in accordance with the Poincar\'e invariance and gave good agreement with
 the
spectrum of the light-quark mesons lying on the leading Regge trajectory. The
next approximation was to take into account the masses and the spins of the 
quarks attached
to the ends of the string. This has been done in [4-16] with various
assumptions. 

The distinctive features of the present approach as compared with those of
refs[4-16] are the consistent treatment
of the quark spins and the canonical quantization. The gauge invariant
formalism is used throughout the paper. We also show that there is no radial
motion of the quarks along the rotating straight-line string. This means that
the daughter meson states correspond to higher modes of the string 
(vibrations). 

 The advantage of the present approach as compared with the potential models
([17] for example) is relativistic invariance (in [17] it is only approximate)
and use of current quark masses (in [17] constituent quark masses are used).
The disadvantage of the present paper is restriction to the leading Regge
trajectories, i.e.,in the potential model language, to the lowest radial
excitations. 

 We consider the Nambu-Goto 
straight-line string with point-like massive spinning quarks attached to its 
ends. This is an extended relativistic object [18,19] called rotator for which 
the explicitly 
relativistic description introduces auxiliary  variables resulting in a 
symmetry
of the rotator Lagrangian. The Hamiltonian of the rotator is given by an 
implicit 
function which can be calculated numerically. For important particular cases
( light or heavy quarks) series expansions for the Hamiltonian are obtained.

The quark spins are described by anticommuting spin variables obeying
constraints [20]. Special care has been taken to ensure conservation of these
constraints [21,22].

Canonical quantization of this system preserving the Poincar\'e invariance 
yields meson states with different spins and parities lying on  Regge 
trajectories which depend on the quark masses. The 16-component wave function 
of a composite meson satisfies two Dirac equations and a spectral condition
which can be compared with the experimental mass spectrum.

The spectral condition contains a contribution of the universal string
confining mechanism together with nonstring short-range contribution which 
is treated phenomenologically. The dominant part of the short-range
contribution do not depend on the meson spin $J$ and its decreasing with $J$
part is seen only in low-$J$ quarkonia.  The string contribution dominates
when at least one quark is light and grows with the meson spin. On the other
hand, it is near threshold for low-spin heavy-quark mesons. The string
contribution to the $\Upsilon (1S)$ mass is about 200 MeV and to the
$\chi_{b2}(1P)$ mass -- 350 MeV.

So, the present approach in its simple form is applicable to mesons containing 
at least one light quark where the nonrelativistic potential models are not
applicable. For heavy quarkonia
the string mechanism should be supplemented with other small ( compared to
heavy-quark masses ) contributions to
account for the fine structure of the levels.
 
We compare the model with experiment for the trajectories  with 
$P = C = (-1)^{J}$ and lowest states having $J^{PC} = 1^{- -}$. For these
trajectories mesons with highest spins were observed and  mixing with
other trajectories is negligible.

This comparison with meson masses allows to estimate the current $s$-, $c$-
and $b$-quark masses assuming that the current $u$- and $d$-quark masses are 
zero within error bars. We then use the chiral $SU_3$ model [23] to estimate
the $u$- and $d$-quark masses through the $s$- quark mass to check the
consistency of the calculations. 

 To check the model we have used the obtained quark masses to calculate 
the masses of mesons not used in the input. We compare the predicted masses
with experiment and with the results of the potential model [17] and discuss a
possible interpretation of the gluon string in terms of the potential model.

So, let us consider a simplest extended relativistic object -- a straight-line
\be
x(\tau, \sigma)=r(\tau)+f(\tau, \sigma)q(\tau),
\ee
where $r$ is a 4-vector corresponding to a point on the straight-line, $q$ is
an affine 4-vector of its direction, $f$ is a scalar monotonic function of
$\sigma$ labelling points on the line and $\tau$ is a scalar evolution 
parameter. We shall not fix the coordinates $f_i(\tau)=f(\tau,\sigma_i(\tau))$
of the end points of the string considering them as dynamical variables to be
determined from extremum of an action. Then the explicit Poincar\'e covariance
of (1) introduces superfluous variables not necessary for description of the
straight-line as a physical object, so that  theory in terms of (1) must be
invariant under a group of three sets of $\tau$-dependent transformations
( gauge transformations )

1)shift of $r$ along $q$:
\be
r\rightarrow r+f(\tau)q,
\ee

2) multiplication of $q$ by an arbitrary scalar function:
\be
q\rightarrow g(\tau)q,
\ee
 
3) reparametrization of $\tau$, what means that the Lagrangian must satisfy
the condition
\be
{\cal L}( h(\tau)\dot z,h(\tau) ( h(\tau)\dot z)\dot {\mbox{}}~)=
h(\tau){\cal L}(\dot z,\ddot z),
\ee
where $\dot z$ and $\ddot z$ mean every $\tau$-derivative in the 
Lagrangian.

This symmetry implies that the phase-space variables of our system obey three
constraints which are in involution with respect to their Poisson brackets; the
canonical Hamiltonian is zero and the total Hamiltonian is a linear combination
of the constraint functions.

Invariants of a symmetry play an important role in description of a 
symmetric system. In our case they are orthonormal vectors along  line
direction, velocity of the line rotation and velocity of its movement as a 
whole
\be
n=(- q^2)^{-1/2}q, ~~ v^1=b^{-1}\dot n, ~~ v^0=(\dot r^2_{\bot})^{-1/2}
\dot r_{\bot},
\ee
where
\be
 b=(-\dot n^2)^{1/2}
\ee
and
\be
\dot r^k_{\bot}=(g^{kl}+n^k n^l +v^{1k} v^{1l})\dot r_l.
\ee
The angular velocity $b$ is invariant under (2) and (3) and transforms as the
Lagrangian under (4). The scalar invariant of the symmetry is
\be
l=b^{-1}(\dot r^2_{\bot} )^{1/2}.
\ee
We shall label points on the string with respect to the instant center of its
rotation $z$
\be
(-q^2)^{1/2} f=z+y,
\ee
\be
z=b^{-1}\dot r v^1
\ee
(velocity of the point $r+zn$, orthogonal to $q$, is orthogonal to $v^1$ ).
The length of the rotator at fixed $\tau$ is $ | y_2-y_1|$. From 
$\dot x^2_i\geq 0$ it follows that $|y_i|\leq l$.
 
We shall take quark spins into account later on. Without quark spins
the Lagrangian of our model is a sum of the Nambu-Goto Lagrangian for an open
string with a string-tension parameter $a$ and two Lagrangians for free 
point-like particles with masses $m_1$ and
$m_2$ and velocities of the ends of the string
\be
{\cal L} = -a \int_{\sigma_1}^{\sigma_2} g^{1/2} d\sigma -
\sum_{i} m_i ( \dot x^2_i)^{1/2} ,
\ee
where $g=(\dot xx')^2-\dot x^2 x'^2$ is minus determinant of the induced 
metric of the string worldsheet and $\dot x_i=dx(\tau,\sigma_i(\tau))/d\tau,~~ 
i=1,2$ are velocities of the string ends. Using the notations 
introduced above we can rewrite (11) for the straight-line string (1,9) in 
the form
\be
{\cal L} = -b F,
\ee
where $F$ is a gauge and Poincar\'e invariant function
\be
F=a\int_{y_1}^{y_2}(l^2-x^2)^{1/2}dx + \sum m_i(l^2 - y^2_i -
w^2_i)^{1/2} ,
\ee
\be
w_i = b^{-1}(\dot y_i + \dot z - \dot rn).
\ee
We shall consider the case when
\be
b \neq 0
\ee
(this is a gauge invariant condition). Then we must consider $w_i$ as
independent variables and the stationary condition with respect to them yields
\be
w_i = 0.
\ee
The other way to obtain this result [16] is to consider the Euler-Lagrange
equations following directly from (11) which  give for the straight-line string 
\be
 \dot y = 0, ~~ \dot z - \dot r n  = 0.
\ee
Eq.(16) follows from here by continuity.

We conclude that for our model
\be
F =a\int_{y_1}^{y_2}(l^2 -x^2)^{1/2}dx +\sum m_i(l^2 -y^2_i)^{1/2}
\ee
with $y_i$ satisfying the stationary condition
\be
\partial F /\partial y_i = 0,
\ee
or
\be
(-1)^i y_i =(l^2 +(m_i/2a)^2)^{1/2} -(m_i/2a).
\ee

Calculating the momenta $p$ and $\pi$ canonically conjugate to $r$ and $q$
\be
p =-\partial {\cal L}/\partial\dot r,~~\pi =-\partial {\cal L}/\partial\dot q
\ee
we get three constraints $\phi_i =0,~~i =1,2,3$ where the constraint functions
are
\be
\phi_1 =pq,~~~~\phi_2 =\pi q,
\ee
\be
\phi_3 =L -K.
\ee
Here
\be
L =((q^2 -(qp)^2/p^2)\pi^2)^{1/2}
\ee
is the magnitude of the conserved orbital spin 
\be
L_{\mu} =\epsilon_{\mu\nu\rho\sigma}p^{\nu} M^{\rho\sigma}/2m
\ee
where 
\be
M^{\mu\nu} =r^{[\mu} p^{\nu]} +q^{[\mu}\pi^{\nu]}
\ee
is the angular momentum tensor. $K$ is a function of $m =(p^2)^{1/2}$, 
implicitly given by the equations
\be
K =lm -F,
\ee
\be
\partial F/ \partial l =m.
\ee

The rotator Hamiltonian is a linear combination of the constraint functions
\be
H =\sum_{i=1,2,3} c_i\phi_i.
\ee
It determines the dynamical equations for any variable $X$
\be
\dot X =\{ X,H\},
\ee
$\phi_i =0$ after calculating the brackets and the non-zero Poisson brackets 
are
\be
\{ p^{\mu}, r^{\nu}\} =\{ q^{\mu}, \pi^{\nu}\} =g^{\mu\nu}.
\ee

We can choose gauge conditions to fix $c_{1,2}=0$ in (29):
\be
p\pi = 0,~~~q^2 + 1 = 0.
\ee
To obtain the Poisson brackets in this gauge we introduce new variables having
vanishing brackets with the constraints (22) and (32)
\be
p,~~r_0 =r +((p\pi)q - (pq)\pi)/p^2,~~v =(-q_p^2)^{-1/2}q_p,~~L
\ee
$(q_p^{\mu} =(g^{\mu\nu} -p^{\mu}p^{\nu}/p^2)q_{\nu})$. 
To have zero brackets of the external coordinate
of the rotation center $r_0$ with the internal coordinates $v$ and $L$ we 
use four orthonormal vectors $e_{\alpha}, \alpha =0,1,2,3$
\be
e_0 =p/m,~~~e_{\alpha} e_{\beta} =g_{\alpha\beta}
\ee
and introduce new variables
\be
n^a =-e_a v,~~~L^a =-e_a L,~~~a =1,2,3,
\ee
\be
z =r_0 +\frac{1}{2}\epsilon_{abc} e_{a\nu}
\frac{\partial e^{\nu}_b}{\partial p} L^c.
\ee
The non-zero Poisson brackets of the new variables are
\be
\{p^k, z^l\} =g^{kl},~~\{L^a, L^b\} =\epsilon_{abc}L^c,~~\{L^a, n^b\} =
\epsilon_{abc} n^c.
\ee
The constraint function $\phi_3$ now takes the form
\be
\phi_3 =((L^a)^2)^{1/2} -K(m)   
\ee
and the solution of the dynamical equations (30) can be easily obtained to be
\be
z =z_0 +lVp/m,
\ee
\be
n =n_0\cos V -n_1\sin V,
\ee
\be
V =\int c_3 d\tau.
\ee
From (39) the laboratory time of the rotation center
\be
t =z^0 -z^0_0 =lVp^0/m
\ee
and the space coordinates of this point
\be
z^a =z^a_0 +p^a t/p^0
\ee
correspond to its movement in the laboratory with constant velocity $p^a/p^0$.
The direction of the rotator rotates with constant angular velocity
\be
\omega =\frac{m}{p^0 l},
\ee
where $l =l(m)$ from (28).

The canonical quantization can now be performed quite easily. We replace our 
variables by operators and their Poisson brackets (37) by commutators. The
constraint equation now holds for the wave function
\be
[((L^a)^2)^{1/2} -K(m) -a_0 ]\psi =0,
\ee
where in the operator form of (38) we have added a term $a_0$ to account for
nonstring short-range contributions.

Our quantum system is relativistic because the quantization procedure 
transforms the classical Poisson brackets of $p^{\mu}$ and $M^{\nu\sigma}$ 
into commutators without any change in their form, so that the Poincar\'e
algebra is fully preserved.

Quark spins are important especially for small $L$. They were taken into
account in [21,22] where the spinless-particle Lagrangians in eq. (11) were
replaced by those of Berezin and Marinov [20] and a special term was added
to preserve conservation of the spin constraints, with the result that for the
leading Regge trajectories one can simply replace the orbital spin $L$ in (45)
by the total meson spin $J$. This yields
\be
(J(J +1))^{1/2} =K(m) +a_0
\ee
 for the physical eigenstates with fixed dependence of space and 
charge-conjugation parities $P$ and $C$ on $J$.

The function $K(m)$ is given by eqs. (18,20,27,28). We must solve eq.(28) to 
find $l$ as a function of $m$ and put this function
into (27). This can be done numerically for any quark masses.  For important 
particular cases $K$ can be expanded into series. For light quarks
\be
y_i =\pi m_i/m \ll 1
\ee
\be
K(m) =\frac{m^2}{2\pi a}[1 -\frac{4}{3\pi}\sum y_i^{3/2}(1 -\frac{3}{20}y_i) +
\frac{1}{(3\pi)^2}(\sum y_i^{3/2})^2 +O(y_i^{7/2})].
\ee
For heavy quarks
\be
D =m -m_1 -m_2 \ll m_i
\ee
\be
K(m) =\frac {1}{a} (\frac{2}{3}D)^{3/2}\nu _1^{-1/2}(1 +
\frac{7}{36}\frac{\nu_3}{\nu_1^2}D +O((\frac{D}{m_i})^2)),
\ee
\be
\nu_n =\sum m_i^{-n}.
\ee
For light and heavy quarks
\be
d =m -m_2,~~y_1 =\frac{\pi m_1}{2d} \ll 1,~~x_2 =\frac{2d}{\pi m_2}\ll 1,
\ee
\ba
K(m) &=&\frac{d^2}{\pi a}
\Bigl [1 -\frac{8}{3\pi}y_1^{3/2} -\frac{2}{\pi}x_2 +
\frac{9}{\pi^2}x_2^2 -(\frac{54}{\pi^3} -\frac{7}{6\pi}) x_2^3 +
(\frac{378}{\pi^4} -\frac{35}{2\pi^2}) x_2^4+\Bigr. \nonumber \\ 
&+& O(y_1^{5/2}) +O(y_1^{3/2} x_2) +
O(x_2^5)\Bigl.\Bigr].
\ea
We see that the slope of the trajectory for mesons formed by a heavy and a
light quark (antiquark) is twice as big as for light-quark mesons.

The term $a_0$ in (46) can in general depend on $J$, but it can not grow with
$J$. An analysis of Coulomb-like short-range interaction suggests the 
following dependence of $a_0$ on $J$ (or on $m$, what is practically the same
when eq.(46) is fulfilled)
\be
a_0 = A + \Bigl (\frac{16m_1 m_2}{(m_1 + m_2)m(2J + 1)^2} \Bigr )^2 B,
\ee
where $A$ and $B$ do not depend on $J$. In all cases considered below the
first term in (54) dominates, so the precise form of the second term is not
important for our conclusions. As a first approximation one could neglect
the second term and to get the quark masses within error bars following
from comparison with experiment. On the other hand the second term allows
one to get good agreement with the experimental heavy-quarkonia spectrum.
The errors in the quark masses in this case formally reduce and to estimate
their values one have to go outside of the model and to analyse the interaction
between mesons and their decay channels. An approximate analysis of this 
problem was performed in ref.[17] with the result that the error in the heavy-
quark meson masses is about 10 Mev. We tentatively take this value as an
error in the heavy-quark masses deduced from a precise fit to experimental 
meson masses with the help of the second term in (54).

Assuming $B$ in (54) to be of order 1 we see that the second term in (54) is
negligible when one or both quarks are light. It is negligible also for
the $s\bar s$-mesons below.

We shall apply Eq.(46) to the leading trajectories with $P = C = (-1)^J$ and 
the lowest states having $J^{PC} = 1^{- -}$. Estimates show they do not mix 
with other trajectories with the same $J^{PC}$ having much heavier states.

Applying eq.(46) to the leading $\rho$ - and $K^{\star}$-trajectories we have
\be
K(m_{\rho J}) =K(m_{K^{\star}J}),
\ee
or, neglecting the $u$- and $d$-quark masses
\be
\frac{m_s}{m_{K^{\star}J}} =\frac{1}{\pi}z_J^{2/3}(1 +\frac{1}{10}z_J^{2/3} +
\frac{1}{18\pi}z_J +O(z_J^{4/3})),
\ee
\be
z_J =\frac{3\pi}{4}(1 -\frac{m^2_{\rho J}}{m^2_{K^{\star}J}}).
\ee
The error from neglecting the $u$- and $d$-quark masses can be estimated
from the $\omega$- and $\rho$-mass difference to be 1.8\%.
Using experimental data for the meson masses from [24] we obtain the
corresponding values for the strange quark mass shown in Table 1.
The error in the average $m_s$ 
corresponds to the accuracy of calculations and, partly, to the accuracy of 
the model.

 We get the following values for the other model parameters
\be
a =.176,~~~ 2\pi a \equiv \alpha'^{-1} =1.11 ~GeV^2,
\ee
\be
a_0 = A = .88.
\ee
The parameter (59) is the same for the light and the strange quarks and 
corresponds to the intercept parameter (of $J$ with the $K =0$ axis) 
$J_0 =.51$.

Knowing the strange-quark mass we can estimate the light-quark masses from
the linear approximation of the chiral $SU_3$ model [23]
\be
m_u/m_d = 0.554 \pm 0.002,~~~m_s/m_d = 20.13 \pm 0.03.
\ee
Using here $m_s$ from Table 1 we get (in Mev)
\be
m_u = 6.2 \pm 0.2,~~~~m_d = 11.1 \pm 0.4.
\ee
We see that neglecting these masses in the above calculations does not 
introduce any noticeable error. 

To check these results we can use them to calculate masses  of mesons
consisting of $s\bar s$, Table 2. They are in good agreement with the
experimental values.

To obtain the $c$-quark mass we consider eqs. (46),(54) for the $D^{\star}$
and $D^{\star}_2$-mesons. The second term in (54) is negligible and
\be
\sqrt{2} = K(D^{\star}) + A(c),~~\sqrt{6} = K(D^{\star}_2) + A(c),
\ee
what allows us to calculate the $c$-quark mass through those of $D^{\star}$
and $D^{\star}_2$ (Table I) and to estimate $A(c)$:
\be
A(c) = .90.
\ee
We see that it is close to the constant $A$ for the light quarks (59). To
describe this closeness let us remark that the shift 0.02 in $a_0$ yields
the shift from $- 10$ to $- 17$ Mev in the vector-meson masses. This shift
decreases for higher meson spins.

Application of eqs.(46),(54) to the $c \bar c$-mesons $J/ \psi$ and
$\chi_{c2}(1P)$ gives the constants
\be
A(c \bar c) = 0.90,
\ee
what coincides with (63) , and
\be
B(c \bar c) = 1.43.
\ee

For the $b$-quark we can not carry out a similar analysis because the mass
of $B^{\star}_2$ is not known. To get an estimate of the $b$-quark mass we
have to rely on an assumption. The safest assumption seems to be
\be
A(b) = A(b \bar b)
\ee
similar to the case of the $c$-quark (63),(64). Using the masses of 
$B^{\star}$,$\Upsilon(1S)$ and $\chi_{b2}(1P)$-mesons we get the $b$-quark
mass in Table I   and
\be
A(b) = A(b \bar b) = 0.77,
\ee
\be
B(b \bar b) = 3.14
\ee
Experimental measurement of the $B^{\star}_2$ mass is important for checking 
the assumption (66).

Now we can calculate masses of other mesons belonging to our trajectory. Some
of them are presented in Table 2, together with experimental data available
and predictions of the potential model of ref.[17]. This model is based upon
linear rising potential, Coulomb-like short-range potential from perturbative
QCD, approximate relativistic corrections and constituent quark masses among
other parameters.

It is tempting to conclude from Table 2 that the present model slightly better
agrees with the data and that future precise measurements might distinguish
both models. But far more impressive is the similarity of the results of
apparently quite different calculations. This similarity confirms the main
physical motivation for considering the gluon string, namely, the string
describes two separate mechanisms of the potential approach, confining 
potential and the constituent quark masses.

In conclusion, let us discuss the relation between quark masses in this model
and in QCD. The present model is a quantum mechanical model of free 
quarks bound in mesons. Since  it agrees with experimental data it is 
reasonable to assume that the quark masses of this model are the current quark 
masses entering as parameters into the QCD Lagrangian when one uses the 
on-mass-shell perturbative renormalization procedure summed to all orders.

It would be interesting to check the obtained values of the current quark
masses in other applications.

The author is grateful to V.A.Petrov, Yu.F.Pirogov and A.V.Razumov for
discussions and to Prof. A. D. Krisch for the kind hospitality at the
University of Michigan where this work was finished.

\begin{table}[htb]
\caption{The input meson masses and the predicted current quark masses in the 
present model.}
\begin{tabular}{|c|c|l|l|}
Meson  &   Input meson       & \multicolumn{2}{c|}{Quark masses in MeV} \\
\cline{3-4}
spin $J$ &  masses[24]       &           Calculated in  &  Other estimates[24]\\
         &                   &         the present model&                  \\
\hline
  1  &    $\rho,~K^{\star}$   &          $m_s =220 \pm 4$ &             \\
  2  &    $a_2,~K_2^{\star}$  &          $m_s =234 \pm 4$ &		\\
  3  &    $\rho_3,~K_3^{\star}$ &        $m_s =204 \pm 18$ &            \\
     &    average              &         $m_s =227 \pm 5$ & $m_s$ =100 to 300 \\
\hline \vrule width0pt height11pt depth0pt
  1  & $D^{\star}, D_2^{\star}$ &    $m_c =1440 \pm 10$ & $m_c$=1.0 to 1.6 GeV\\
  1  & $\Upsilon,~B^{\star},\chi_{b2}$ & $m_b =4715 \pm 20$  & $m_b$=4.1 to 4.5 GeV\\
\end{tabular}
\end{table}

\begin{table}[htb]
\caption{ The model predictions for  meson masses (in Mev) and comparison with 
the potential model predictions of ref.[17] ($q$ stands for $u$ or $d$). }
\begin{tabular}{|c|c|l|l|c|}
Quark   &  Meson  & Present         &  Experimental    & Potential\\
content & spin $J^{PC}$ & model     & values           & model [17]\\		
\hline
$q\bar q$ & $2^{++}$ &  1317          & $1318.1 \pm 0.7$   &  1310\\
        & $3^{--}$ &  1690          & $1691 \pm 5$       &  1680\\
        & $4^{++}$ &  1993          &                &  2010\\
        & $5^{--}$ &  2255          &                &  2300\\
\hline
$q\bar s$ & $4^{+}$  &  2080          & $2045 \pm 9$       &  2110\\
\hline
$s\bar s$ & $1^{--}$ &  1019          & $1019.413 \pm 0.008$ & 1020\\
      & $2^{++}$ &  1520          & $1525 \pm 5$      &  1530\\
      & $3^{--}$ &  1873          & $1854\pm 7$       &  1900\\
      & $4^{++}$ &  2160          &                &  2200\\
\hline
$c\bar q$ & $3^{-}$  &  2780          &               &  2830\\
\hline
$c\bar s$ & $1^{-}$  &  2134          & $2112.4 \pm 0.7$   &  2130\\
        & $2^{+}$  &  2561          & $2573.5 \pm 1.7$   &  2590\\
        & $3^{-}$  &  2870          &                &  2920\\
\hline
$c\bar c$ & $3^{--}$ &  3830          &                &  3850\\
\hline \vrule width0pt height11pt depth0pt
$b\bar q$ & $2^{+}$  &  5720          &                &  5800\\
\hline
$b\bar s$ & $1^{-}$  &  5430          &                &  5450\\
\hline
$b\bar c$ & $1^{-}$  &  6410          &                &  6340\\
\hline 
$b\bar b$ & $3^{--}$ &  10110         &                &  10160\\
\end{tabular}\end{table}

\end{document}